\documentclass[aps,pre,twocolumn,amsmath,amssymb,nofootinbib,superscriptaddress,floatfix]{revtex4-1}

\usepackage{amsmath,amsfonts,amssymb}
\usepackage{amssymb}
\usepackage{amsthm}
\usepackage[dvipdf]{color}
\usepackage{graphicx}
\usepackage{dcolumn}
\usepackage{bm} 
\usepackage{ulem}
\usepackage{cancel}

\begin{document}
\title{Floquet Chern Vector Topological Insulators in Three Dimensions}
	
\author{Fangyuan Ma}
\thanks{These authors contribute equally to this work.}
\affiliation{Key Lab of Advanced Optoelectronic Quantum Architecture and Measurement (MOE), School of Physics, Beijing Institute of Technology, Beijing, 100081, China}

\author{Junrong Feng}
\thanks{These authors contribute equally to this work.}
\affiliation{Key Lab of Advanced Optoelectronic Quantum Architecture and Measurement (MOE), School of Physics, Beijing Institute of Technology, Beijing, 100081, China}

\author{Feng Li}
\email{phlifeng@bit.edu.cn}
\affiliation{Key Lab of Advanced Optoelectronic Quantum Architecture and Measurement (MOE), School of Physics, Beijing Institute of Technology, Beijing, 100081, China}

\author{Ying Wu}
\email{yingwu@njust.edu.cn}
\affiliation{School of Physics, Nanjing University of Science and Technology, Nanjing 210094, China}

\author{Di Zhou}
\email{dizhou@bit.edu.cn}
\affiliation{Key Lab of Advanced Optoelectronic Quantum Architecture and Measurement (MOE), School of Physics, Beijing Institute of Technology, Beijing, 100081, China}

\begin{abstract}
We theoretically and numerically investigate Chern vector insulators and topological surface states in a three-dimensional lattice, based on phase-delayed temporal-periodic interactions within the tight-binding model. These Floquet interactions break time-reversal symmetry, effectively inducing a gauge field analogous to magnetic flux. This gauge field results in Chern numbers in all spatial dimensions, collectively forming the Chern vector. This vector characterizes the topological phases and signifies the emergence of robust surface states. Numerically, we observe these states propagating unidirectionally without backscattering on all open surfaces of the three-dimensional system. Our work paves the way for breaking time-reversal symmetry and realizing three-dimensional Chern vector topological insulators using temporal-periodic Floquet techniques.
\end{abstract}
	
\maketitle
	
\section{Introduction}
Topological insulators possess robust edge states protected by nontrivial band topology~\cite{PhysRevLett.98.106803, Kane2005PRL, RevModPhys.82.1959, PhysRevB.101.115413, PhysRevB.108.125147}, making them highly useful in spintronics and quantum computation~\cite{Zhang2011RMP, Kane2010RMP, PhysRevX.1.021014, ovchinnikov2022topological}. In two-dimensional systems, Chern insulators host chiral edge states that propagate unidirectionally without backscattering~\cite{PhysRevLett.61.2015, PhysRevLett.102.107603, PhysRevLett.109.186805}. These robust states are protected by the Chern number, which reflects the topological nature of the band structure defined in reciprocal space. Originally constructed from quantum topological band theory, these non-trivial properties have been demonstrated to ubiquitously arise in modern classical structures such as photonic~\cite{devescovi2021cubic, PhysRevB.109.L140104}, mechanical~\cite{susstrunk2015observation, nash2015topological, PhysRevLett.123.034301}, fluid~\cite{khanikaev2015topologically, PhysRevLett.122.128001, souslov2017topological, dasbiswas2018topological}, plasmonic~\cite{fu2021topological, PhysRevResearch.6.023273}, and electrical circuit~\cite{PhysRevB.107.L201101} systems, offering great opportunities for the novel control of classical waves~\cite{sone2024nonlinearity}.
 
Recently, the concepts of the Chern number and the two-dimensional Chern insulator have been extended to three-dimensional systems~\cite{linyun2024acoustic}. These systems' band structures exhibit three Chern numbers, which together constitute the Chern vector. This topologically protected vector indicates that topological states can emerge and propagate not only along the one-dimensional edges of two-dimensional materials but also on the two-dimensional surfaces of three-dimensional structures. This intriguing physics has been experimentally demonstrated in photonic metamaterials~\cite{liu2022topological} incorporating gyromagnetic elements that break time-reversal symmetry. Consequently, Chern vector topological insulators represent the three-dimensional extension of two-dimensional Chern insulators.

Breaking time-reversal symmetry is difficult to achieve in classical structures. For instance, the experimental observation of Chern vector photonic states requires a strong and spatially uniform magnetic field \cite{liu2022topological, hu2024observation}, which is challenging. Additionally, the mechanical analog of the magnetic field, often realized through the Coriolis force, necessitates high-speed rotation \cite{PhysRevB.96.064106}. 

Among the various methods of breaking time-reversal symmetry in classical structures, time-modulation stands out as a promising alternative. 
Time-modulated interactions induce Floquet topological insulators, which have been experimentally realized over the past decade~\cite{PhysRevB.109.125303, kawai2002parametrically, liu2024tuning, fan2021floquet, fleury2016floquet, PhysRevLett.123.016806, gomez2013floquet, han2020classification, wen2018floquet, PhysRevE.110.015003}. This technique, often involving active elements that require energy input, is easier to control.  
Previous works have utilized temporal-periodic elements to break time-reversal symmetry in classical structures and realize two-dimensional Chern insulators. Given the potential of temporal-periodic interactions in breaking time-reversal symmetry, it is compelling to explore whether time-modulated models can be applied to classical structures to realize topological insulators characterized by Chern vector topological invariants and their corresponding novel physical properties.

In this work, based on a tight-binding model, we propose a theoretical approach using a set of ``phase-delayed" temporal-periodic interactions. 
Our model is based on the network connectivity of a modified stacked kagome lattice~\cite{baardink2018localizing} in three dimensions. From a geometric perspective, lattice edges can form closed loops by connecting the head to the tail of the bonds, creating enclosed areas. We ensure that each bond has a constant delayed phase compared to the next bond in this loop, producing a non-zero effective gauge field analogous to the magnetic flux in the Haldane model, effectively breaking time-reversal symmetry. 
Using Bloch-Floquet analysis~\cite{eckardt2015high, parker2023standing, melkani2024space, Chaunsali2019PRB} of the initial time-dependent model, we obtain a Floquet Hamiltonian. With non-zero projections of the time-reversal-breaking gauge field in all three spatial directions, the band structure of the Floquet Hamiltonian acquires topological Chern numbers in all spatial dimensions, forming a topologically robust Chern vector. We theoretically and numerically demonstrate that topological chiral edge states can propagate without backscattering along all surfaces of the time-dependent model. Our work paves the way for novel designs and realizations of Chern vector topological insulators.



\begin{figure*}
\centering
\includegraphics[width=0.95\textwidth]{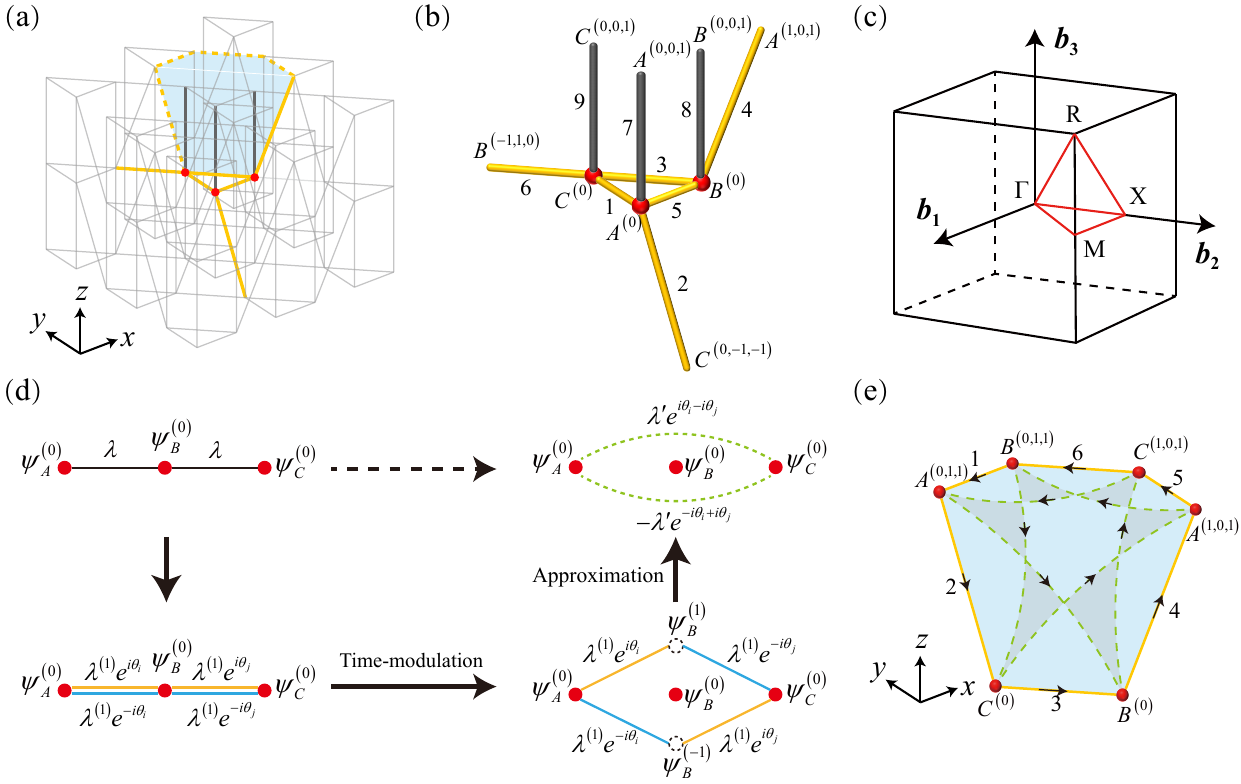}
\caption{Schematic illustration of the tight binding model that is embedded in the three-dimensional modified stacked kagome lattice. (a) and (b) show the unit cell and the convention of site and bond labeling. (c) displays the three-dimensional Brillouin zone with high-symmetry points. (c) illustrates a closed loop that is constructed by a set of temporally-modulated bonds by connecting the bonds from the head to the tail, inducing a non-zero effective magnetic flux in their enclosed area. Arrows denote the complex hopping between connected sites, with a phase of $\pi/2$, effectively inducing a gauge flux of $\pi/2$ for each gray-colored (curvy triangular) area. (d) Schematic illustration of the interaction between higher-order harmonic waves at $B$ and zeroth-order waves at $A$ and $C$, resulting in the mutual long-range (next-nearest-neighbor) interactions between $A$ and $C$ that were not connected in the initial model.  \label{lattice}}
\end{figure*}

This article is structured as follows: In Section \uppercase\expandafter{\romannumeral2}, we introduce a tight-binding model with periodically time-modulated hopping terms, embedded in the three-dimensional structure known as the stacked kagome lattice. In Section \uppercase\expandafter{\romannumeral3}, we employ Bloch-Floquet analysis to simplify the model. In Section \uppercase\expandafter{\romannumeral4}, we consider the high-frequency limit and implement perturbative analysis, absorbing the effects of higher-order harmonic modes into the fundamental harmonics. We derive the gauge field that breaks time-reversal symmetry. In Section \uppercase\expandafter{\romannumeral5}, we compute the topological Chern vector of the Floquet band structure and numerically demonstrate the unidirectional propagation of topological surface states on the two-dimensional surface of a three-dimensional lattice. We further verify the robustness of these surface states against structural defects.

\section{Tight Binding Model with Temporal-Periodic Hopping}

Among various theoretical models, the tight-binding model is widely used to study topological physics in the Su-Schriffer-Heeger model~\cite{Su1979PRL}, the Chern insulator~\cite{PhysRevB.109.195101}, the quantum spin-Hall insulator~\cite{bernevig2006quantum}, and higher-order topological insulators~\cite{PhysRevLett.123.053902}. Originally designed for quantum electronic systems, the tight-binding model also captures the foundations of classical systems~\cite{duncan2020topological}, such as mechanical~\cite{Kane2014NP, Zhou2019PRX, fruchart2020dualities, PhysRevLett.133.106101}, photonic~\cite{tuloup2023breakdown}, phononic~\cite{lee2022piezoelectric}, and electrical systems~\cite{liu2020octupole}.

In this paper, we study a tight-binding model with time-periodically modulated interactions to investigate the realization of Chern vector topological insulators. Our model is based on a three-dimensional lattice, created by stacking two-dimensional kagome lattices and modifying their connectivity in the third dimension to form a modified stacked kagome lattice, as detailed in Fig. \ref{lattice}(a). The unit cell contains three sites at positions $A=(0, 0, 0)$, $B=(1, 0, 0)$, and $C=(0, 1, 0)$. The primitive vectors are $\bm{a}_1=(2, 0, 0)$, $\bm{a}_2=(0, 2, 0)$, and $\bm{a}_3=(0, 0, 2)$. Each unit cell is labeled by three integer-valued indices $n_1$, $n_2$, $n_3$, with its position given by $\bm{r} = n_1 \bm{a}_1 + n_2 \bm{a}_2 + n_3 \bm{a}_3$. We define the reciprocal vectors as $\bm{b}_\alpha = 2\pi\epsilon_{\alpha\beta\gamma}\bm{a}_\beta\times \bm{a}_\gamma/\bm{a}_\alpha\cdot(\bm{a}_\beta\times\bm{a}_\gamma)$ for $\alpha,\beta,\gamma=1,2,3$. We further simplify the notation using the three-component vector index $\bm{n} = (n_1, n_2, n_3)$, so that every node in the unit cell can be denoted as $X(\bm{n})$ for $X = A, B, C$. The in-plane triangle of the stacked kagome lattice is an isosceles right triangle. While the geometric shape of this triangle does not affect the interactions, we chose it for its simplicity. The primitive vector directions are oriented along the $x$ and $y$ axes, significantly simplifying the geometry of this system.

We label the temporally-oscillating interaction bonds (i.e., the edge vectors that connect the sites) as $l_i$ for $i=1, 2, \ldots, 9$. The edge vectors that represent the bonds are defined as follows: 
\begin{eqnarray}\label{edge} 
& {} & l_1 = A(\bm{n})C({n}), \nonumber \\ & {} & l_2 = C(\bm{n})A(\bm{n} - (0, 1, 1)), \nonumber \\ & {} & l_3 = C(\bm{n})B(\bm{n}), \nonumber \\ & {} & l_4 = A(\bm{n} + (1, 0, 1))B(\bm{n}), \nonumber \\ & {} & l_5 = B\bm({n})A(\bm{n}), \nonumber \\ & {} & l_6 = B(\bm{n} + (-1, 1, 0))C(\bm{n}), \nonumber \\ & {} & l_7 = A(\bm{n} + (0, 0, 1))A(\bm{n}), \nonumber \\ & {} & l_8 = B(\bm{n} + (0, 0, 1))B(\bm{n}), \nonumber \\ & {} & l_9 = C(\bm{n} + (0, 0, 1))C(\bm{n}). 
\end{eqnarray}
There are a total of nine bonds in the unit cell. Except for $l_2$ and $l_4$, the edge vectors $l_i$ follow the conventional definition of nearest-neighbor bonds in the regular stacked kagome lattice. Meanwhile, $l_2$ and $l_4$ are modified bonds deviating from this regular connectivity. Specifically, the $l_2$ edge vector denotes the time-periodic interaction between the $C(\bm{n})$-site and the $A(\bm{n} - (0, 1, 1))$-site, while the $l_4$ edge denotes the bond between the $B(\bm{n})$-site and the $A(\bm{n} + (1, 0, 1))$-site.

We denote $\psi_{X(\bm{n})}$ for $X=A,B,C$ as the wave function of site $X$ in the unit cell at position $\bm{r}(\bm{n})$. Furthermore, we use $\psi=(\ldots, \psi_{A(\bm{n})}, \psi_{B(\bm{n})}, \psi_{C(\bm{n})}.\ldots)$ to denote the wave function of the entire lattice. Thus, the field dynamics of the lattice is governed by the time-dependent Schr\"{o}dinger equation:
\begin{eqnarray}\label{EOM}
\mathrm{i}\partial_t \psi(t) = H(t)\psi(t),
\end{eqnarray}
where $\mathrm{i}=\sqrt{-1}$, the Plank's constant is set to $\hbar=1$ for simplicity, and $H(t)$ is the time-dependent Hamiltonian:
\begin{eqnarray}\label{real_space_TB}
 & {} & H(t) 
=m\sum_{X(\bm{n})}\hat{c}_{X(\bm{n})}^\dag c_{X(\bm{n})}\nonumber \\
 & {} & +\sum_{i=1}^6\sum_{l_i =  X'(\bm{n})X(\bm{n})}\lambda_i(t) \left(\hat{c}^\dag_{X'(\bm{n}')}c_{X(\bm{n})}+h.c.\right) \nonumber \\
 & {} & +\sum_{i=7,8,9}\sum_{l_i =  X'(\bm{n})X(\bm{n})}\lambda_z \left(\hat{c}^\dag_{X'(\bm{n}')}c_{X(\bm{n})}+h.c.\right).
\end{eqnarray}
Here, $m$ is the on-site potential, $\lambda_z$ is the static hopping coefficient for the bonds labeled $i=7,8,9$, and $\hat{c}_{X(\bm{n})}^\dag$ and $\hat{c}_{X(\bm{n})}$ are the creation and annihilation operators for particles at sites $X$ of the unit cell labeled $\bm{n}$. The temporal-periodic interactions that connect the sites, namely $\lambda_i(t)$ for $i=1,2,\ldots, 6$, are expressed as 
\begin{eqnarray}\label{Floquet_coup} 
\lambda_i(t)=\lambda^{(0)}+\lambda^{(1)} \cos(\omega_F t+\phi_i),
\end{eqnarray} 
where $\lambda^{(0)}$ and $\lambda^{(1)}$ are real-valued constants that will be specified in Section \uppercase\expandafter{\romannumeral5} for numerical computations. These are temporal-periodic functions with the angular frequency $\omega_F$. Furthermore, each temporal-periodic coupling has a different phase $\phi_i$. For the sake of simplicity, we assume that each bond has a constant delayed phase compared to the next bond, which is 
\begin{equation}\label{phase} 
\phi_{i+1}=\phi_i+\Delta\phi , \quad \Delta\phi = 2\pi/N, \quad (N=6)
\end{equation} 
for the six bonds in a loop. Here, as long as the phase difference, $\Delta\phi = \phi_{i+1} - \phi_i \neq 0, \pi$, the time-dependent hopping interaction is distinct after the reversal of time, naturally breaking time-reversal symmetry. This can be directly observed by noting that under the operator of time-reversal symmetry, the time-dependent coefficient transforms into \begin{equation}\label{TR_breaking1} 
\mathcal{T}\lambda_i(t) = \lambda_i^*(-t) = \lambda^{(0)} + \lambda^{(1)} \cos(\omega_F t - \phi_i). 
\end{equation} 
Here, the time-reversal operator is $\mathcal{T} = (\mathcal{K}, t \to -t)$ in real space, which combines complex conjugation and the reversal of time. Under the operation of time reversal, the phase difference $\Delta\phi$ flips sign from $2\pi/N$ to $-2\pi/N$, corresponding to the advanced phase for each bond compared to the next. This, in turn, leads to a non-zero net gauge flux in the loop enclosed by the bonds, causing the time-reversal breaking nature of the resulting effective Hamiltonian and inducing Chern vectors and chiral topological edge states. We will discuss this in detail in Section \uppercase\expandafter{\romannumeral5}, where the topological Chern vector is introduced.



\section{Bloch-Floquet Analysis}
Having established the time-dependent model, we now employ Bloch-Floquet analysis to simplify the system and theoretically study its properties. First, we convert the system from real space to momentum space. This is achieved by using Fourier transformation on the wave function, described by the relationship $\psi_{\bm{n}} = \sum_{\bm{q}} \psi_{\bm{q}} e^{\mathrm{i} \bm{q} \cdot \bm{r}(\bm{n})}$, where $\bm{q}$ is the three-dimensional wavevector in momentum space, and $\psi_{\bm{q}}$ is the wave function with momentum $\bm{q}$.

Furthermore, we define $k_i = \bm{q} \cdot \bm{a}_i$ for $i = 1, 2, 3$, which characterizes the phase shift of the wave component along the primitive vector $\bm{a}_i$ and resides in the region $-\pi < k_i \le \pi$. As a result, the wave function can be further decomposed as $\psi_{\bm{n}} = \sum_{\bm{k}} \psi_{\bm{k}} e^{\mathrm{i} \bm{k} \cdot \bm{n}}$, where $\bm{k} = (k_1, k_2, k_3)$ is the three-component wavevector in units of the primitive vector. Within Bloch analysis, the Hamiltonian can be written as $H_{\bm{k}}(t)$, allowing the Schr\"{o}dinger equation to be written as the following form: $\mathrm{i}\partial_t \psi_{\bm{k}}(t)=H_{\bm{k}}(t)\psi_{\bm{k}}(t)$. We further simplify this Hamiltonian using a Fourier series in frequency space, writing it in the following format: 

\begin{widetext}

		\begin{align}\label{Fourier_Freq}
			\begin{split}
   & H_{\bm{k}}(t) = H^{(0)}_{\bm{k}} + H^{(1)}_{\bm{k}} e^{+\mathrm{i}\omega_F t} + H^{(-1)}_{\bm{k}} e^{-\mathrm{i}\omega_F t}, \\
				& H^{(0)}_{\bm{k}}=\left( m+2\lambda_{z}\cos k_3 \right)\textbf{I}_{3\times 3}+\lambda^{(0)}\left( \begin{matrix}
					0 & 1+e^{-\mathrm{i}k_1-\mathrm{i}k_3} & 1+{e^{-\mathrm{i}k_2-\mathrm{i}k_3}}  \\
					1+e^{\mathrm{i}k_1+\mathrm{i}k_3} & 0 & 1+e^{\mathrm{i}k_1-\mathrm{i}k_2}  \\
					1+{{e}^{\mathrm{i}{{k}_{2}}+i{{k}_{3}}}} & 1+{{e}^{-\mathrm{i}{{k}_{1}}+\mathrm{i}{{k}_{2}}}} & 0  \\
				\end{matrix} \right), \\ 
				& H^{( \pm 1 )}_{\bm{k}}=\frac{\lambda^{(1)} }{2}\left( \begin{matrix}
					0 & {{e}^{\mp \mathrm{i}{{\phi }_{1}}}}+{{e}^{-\mathrm{i}{{k}_{1}}-\mathrm{i}{{k}_{3}}\mp \mathrm{i}{{\phi }_{4}}}} & {{e}^{\mp \mathrm{i}{{\phi }_{5}}}}+{{e}^{-\mathrm{i}{{k}_{2}}-\mathrm{i}{{k}_{3}}\mp \mathrm{i}{{\phi }_{2}}}}  \\
					{{e}^{\mp \mathrm{i}{{\phi }_{1}}}}+{{e}^{\mathrm{i}{{k}_{1}}+\mathrm{i}{{k}_{3}}\mp \mathrm{i}{{\phi }_{4}}}} & 0 & {{e}^{\mp \mathrm{i}{{\phi }_{3}}}}+{{e}^{\mathrm{i}{{k}_{1}}-i{{k}_{2}}\mp \mathrm{i}{{\phi }_{6}}}}  \\
					{{e}^{\mp \mathrm{i}{{\phi }_{5}}}}+{{e}^{i{{k}_{2}}+i{{k}_{3}}\mp \mathrm{i}{{\phi }_{2}}}} & {{e}^{\mp \mathrm{i}{{\phi }_{3}}}}+{{e}^{-\mathrm{i}{{k}_{1}}+\mathrm{i}{{k}_{2}}\mp \mathrm{i}{{\phi }_{6}}}} & 0  \\
				\end{matrix} \right),
			\end{split}
		\end{align}
	\end{widetext}
where $\textbf{I}_{3\times 3}$ is the $3\times 3$ identity matrix. From Eq. (\ref{Fourier_Freq}), the Hamiltonian consists of a static term and two harmonic components with frequencies of $\pm\omega_F$.


Due to the time-periodic nature of the Hamiltonian, the amplitude of the responding wave function also exhibits a time-periodic feature. The wave function can be expressed as the product of two parts: the first part, $e^{-\mathrm{i}\omega_{\bm{k}} t}$, commonly arises from the static Schr\"{o}dinger problem. The quasi-energy (quasi-eigenvalue) $\omega_{\bm{k}}$ governs the temporal modulation in the phase variable and resembles the eigenvalues in the ordinary Schr\"{o}dinger equation if the Hamiltonian were static in time. The second part arises from the oscillating nature of the temporal-periodic Hamiltonian itself and reflects the oscillating behavior of the wave function's amplitude~\cite{Zhou2022NC, tempelman2021topological}. This part is expressed as a Fourier series in the frequency space: $\sum_n \varphi^{(n)}_{\bm{k}} e^{-\mathrm{i} n \omega_F t}$, where $n$ is an integer that denotes the Fourier series, and $\varphi^{(n)}_{\bm{k}}$ is a $3\times 1$ vectorial coefficient that reflects the $n$th Fourier component of the three-component unit cell wave function. It is intuitive to observe that, in a purely static Schr\"{o}dinger equation, the mode amplitude should not vary in time, and the aforementioned Fourier series would contain only static components. Due to the temporal-periodic nature of the mutual interactions between nearest-neighbor sites, the aforementioned quasi-eigenvalue can differ from the regular eigenvalue in a purely static Hamiltonian. Consequently, the wave function can be expressed in the following form:
\begin{equation}
\psi_{\bm{k}}(t) = e^{-\mathrm{i}{\omega_{\bm{k}} t}}\sum\limits_{n  \in \mathcal{Z}} \varphi^{(n)}_{\bm{k}}{e^{ - \mathrm{i}n \omega_F t}}.
\end{equation}
Using these Fourier components of the wave function, we construct the Floquet wave function via $\varphi_{\bm{k}} = (\ldots, \varphi^{(-2)}_{\bm{k}}, \varphi^{(-1)}_{\bm{k}}, \varphi^{(0)}_{\bm{k}}, \varphi^{(1)}_{\bm{k}}, \varphi^{(2)}_{\bm{k}}, \ldots)$, which is an infinite-dimensional vector that captures all Fourier components in the frequency space of the wave function. Substituting this Floquet wave function back into the Schr\"{o}dinger equation, we arrive at the Floquet eigenvalue equation:
\begin{equation}
	\label{eigen_equ}
\mathcal{H}_{\bm{k}}\varphi_{\bm{k}}=\omega_{\bm{k}}\varphi_{\bm{k}},
\end{equation}
where the Floquet Hamiltonian $\mathcal{H}_{\bm{k}}$ is explicitly expressed as follows, 
\begin{eqnarray}\label{floquet_H}
 & {} & \mathcal{H}_{\bm{k}} = \nonumber \\
 & {} & 
 \left(\begin{array}{ccccc}
\ldots & \ldots & \ldots & \ldots & \ldots \\
\ldots & H^{(0)}_{\bm{k}}-\omega_F & H^{(-1)}_{\bm{k}} & 0 & \ldots \\
\ldots & H^{(1)}_{\bm{k}} & H^{(0)}_{\bm{k}} & H^{(-1)}_{\bm{k}} & \ldots \\
\ldots & 0 & H^{(1)}_{\bm{k}} & H^{(0)}_{\bm{k}}+\omega_F & \ldots \\
\ldots & \ldots & \ldots & \ldots & \ldots \\
\end{array}\right).
\end{eqnarray}
As shown in Eq. (\ref{floquet_H}), the Floquet Hamiltonian is written in terms of block matrices $H_{\bm{k}}^{(0)}$, $H_{\bm{k}}^{(-1)}$, and $H_{\bm{k}}^{(1)}$. This Floquet Hamiltonian is an infinite-dimensional matrix, making it impractical to solve analytically. Therefore, in the next section, we will employ analytical tools to solve this Floquet problem perturbatively by considering the high-frequency limit.

\section{Time-reversal-breaking Hamiltonian and gauge field}

In principle, the Floquet Hamiltonian presented in Eq. (\ref{floquet_H}) is an infinite-dimensional matrix and it is therefore impossible to compute all quasi-energy of this Floquet problem. However, in the high-frequency limit where $\omega_F\gg \max(m, \lambda^{(0)}, \lambda^{(1)}, \lambda_z)$, we can perturbatively compute the energy eigenvalues by expanding the problem with respect to $1/\omega_F$. This is analogous to the idea of rotating wave approximation, in which higher-order harmonics of the wave function are small comparing to the static component wave function.

One widely used method to study the Floquet Hamiltonian perturbatively is to truncate the infinite-dimensional Floquet Hamiltonian to a finite matrix. This involves expressing higher order harmonics in terms of lower order harmonic modes. For instance, all higher order harmonic modes $\varphi_{\bm{k}}^{(n)}$ with $n \geq 2$ are expressed in terms of lower order modes $\varphi_{\bm{k}}^{(n)}$ with $n = 0, 1$. The resulting matrix Hamiltonian, known as the truncated Floquet Hamiltonian within the first order harmonics, is a $9 \times 9$ matrix Hamiltonian. This matrix is mathematically represented as a $3 \times 3$ block matrix:
\begin{eqnarray}\label{floquet_H2}
\mathcal{H}_{\bm{k}}^{(1)} = 
 \left(\begin{array}{ccc}
H^{(0)}_{\bm{k}}-\omega_F & H^{(-1)}_{\bm{k}} & 0 \\
H^{(1)}_{\bm{k}} & H^{(0)}_{\bm{k}} & H^{(-1)}_{\bm{k}}\\
0 & H^{(1)}_{\bm{k}} & H^{(0)}_{\bm{k}}+\omega_F \\
\end{array}\right).
\end{eqnarray}
Due to the time-modulated hopping interactions breaking time-reversal symmetry, this Hamiltonian inherently breaks time-reversal symmetry, resulting in the non-zero Chern vector discussed in Section \uppercase\expandafter{\romannumeral5}.

We further employ a perturbation method, specifically the Floquet-Magnus expansion~\cite{eckardt2015high}, to reduce the infinite-dimensional Floquet Hamiltonian to a $3 \times 3$ matrix. This is achieved by expressing higher-order harmonics of the wave function, $\varphi_{\bm{k}}^{(n)}$ with $n \geq 1$, in terms of the lowest-order (zeroth-order) harmonic component $\varphi_{\bm{k}}^{(0)}$, which is the static component of the wave function's Fourier series. By expressing all high-order harmonics in terms of the lowest-order static component, we obtain a Floquet problem that is zeroth order in the Fourier components of the wave functions:
\begin{eqnarray}\label{floquet_H0}
\mathcal{H}_{\bm{k}}^{(0)}\varphi_{\bm{k}}^{(0)} = \epsilon_{\bm{k}}\varphi_{\bm{k}}^{(0)},
\end{eqnarray}
where $\varphi_{\bm{k}}^{(0)}$ is the lowest-order (static) component of the wave function, $\epsilon_{\bm{k}}$ is the eigenvalue of the truncated Floquet Hamiltonian, and the $3\times 3$ truncated Floquet Hamiltonian is mathematically expressed as follows according to the Floquet-Magnus expansion:
\begin{eqnarray}\label{floquet_H0_2}
\mathcal{H}_{\bm{k}}^{(0)} = H_{\bm{k}}^{(0)}+\frac{1}{\omega_F}[H_{\bm{k}}^{(-1)},H_{\bm{k}}^{(1)}]+\mathcal{O}\left(\frac{1}{\omega_F^2}\right).
\end{eqnarray}
Here, $H_{\bm{k}}^{(0)}$ is the Hamiltonian that arises purely from the static part of the mutual interactions (see Eqs. (\ref{Fourier_Freq})). This part yields time reversal symmetry, 
\begin{eqnarray}
\mathcal{T} H_{\bm{k}}^{(0)}\mathcal{T}^{-1} = H_{-\bm{k}}^{(0)} ,
\end{eqnarray}
where the time-reversal operator in reciprocal space reads $\mathcal{T}=\mathcal{K}$, where $\mathcal{K}$ is the operator of complex conjugation. On the other hand, we find $\mathcal{T}H_{\bm{k}}^{(\pm1)}\mathcal{T}^{-1}=H_{-\bm{k}}^{(\mp1)}$, leading to the result
\begin{eqnarray}
\mathcal{T} [H_{\bm{k}}^{(-1)},H_{\bm{k}}^{(1)}]\mathcal{T}^{-1}=-[H_{-\bm{k}}^{(-1)},H_{-\bm{k}}^{(1)}].
\end{eqnarray}
This relationship states that the perturbing Hamiltonian that arises from the time modulation in the bonds can break time-reversal symmetry. We further rewrite this perturbing term as following, 

 \begin{widetext}

		\begin{equation}\label{floquet_H0_3}
			\begin{split}
   & [H_{\bm{k}}^{(-1)},H_{\bm{k}}^{(1)}] = H_{\bm{k}}^{\rm NN}+H_{\bm{k}}^{\rm NNN}, \\
				& H_{\bm{k}}^{\rm NN} = \frac{1}{2}\lambda^{(1)2} e^{\mathrm{i}\left(\frac{\pi}{2}+\pi{\,\rm sgn}(\sin\Delta\phi)\right)}|\sin\Delta\phi|\left( \begin{matrix}
					0 & 1+{{e}^{-\mathrm{i}{{k}_{1}}-\mathrm{i}{{k}_{3}}}} & -1-{{e}^{-\mathrm{i}{{k}_{2}}-\mathrm{i}{{k}_{3}}}}  \\
					-1-{{e}^{\mathrm{i}{{k}_{1}}+\mathrm{i}{{k}_{3}}}} & 0 & 1+{{e}^{\mathrm{i}{{k}_{1}}-\mathrm{i}{{k}_{2}}}}  \\
					1+{{e}^{\mathrm{i}{{k}_{2}}+\mathrm{i}{{k}_{3}}}} & -1-{{e}^{-\mathrm{i}{{k}_{1}}+\mathrm{i}{{k}_{2}}}} & 0  \\
				\end{matrix} \right), \\ 
				& H_{\bm{k}}^{\rm NNN}= \frac{1}{2}\lambda^{(1)2} e^{\mathrm{i}\left(\frac{\pi}{2}+\pi{\,\rm sgn}(\sin\Delta\phi)\right)}|\sin\Delta\phi|\left( \begin{matrix}
					0 & -{{e}^{-\mathrm{i}{{k}_{1}}+\mathrm{i}{{k}_{2}}}}-{{e}^{-\mathrm{i}{{k}_{2}}-\mathrm{i}{{k}_{3}}}} & {{e}^{\mathrm{i}{{k}_{1}}-\mathrm{i}{{k}_{2}}}}+{{e}^{-\mathrm{i}{{k}_{1}}-\mathrm{i}{{k}_{3}}}}  \\
					{{e}^{\mathrm{i}{{k}_{1}}-\mathrm{i}{{k}_{2}}}}+{{e}^{\mathrm{i}{{k}_{2}}+\mathrm{i}{{k}_{3}}}} & 0 & -{{e}^{-\mathrm{i}{{k}_{2}}-\mathrm{i}{{k}_{3}}}}-{{e}^{\mathrm{i}{{k}_{1}}+\mathrm{i}{{k}_{3}}}}  \\
					-{{e}^{-\mathrm{i}{{k}_{1}}+\mathrm{i}{{k}_{2}}}}-{{e}^{\mathrm{i}{{k}_{1}}+\mathrm{i}{{k}_{3}}}} & {{e}^{\mathrm{i}{{k}_{2}}+\mathrm{i}{{k}_{3}}}}+{{e}^{-\mathrm{i}{{k}_{1}}-\mathrm{i}{{k}_{3}}}} & 0  \\
				\end{matrix} \right),
			\end{split}
		\end{equation}

	\end{widetext}
where $H_{\bm{k}}^{\rm NN}$ and $H_{\bm{k}}^{\rm NNN}$ denote nearest-neighbor hopping and next-nearest-neighbor hopping that break time-reversal symmetry, respectively. The phase delay of each bond compared to the next is characterized by $\Delta\phi = \frac{2\pi}{N}$ with $N = 6$.

It is worth emphasizing that in the initial time-dependent Hamiltonian, as shown in Eq. (\ref{real_space_TB}), only nearest-neighbor time-dependent hopping exists, while next-nearest-neighbor interactions do not. However, due to the temporal-periodic nature of the mutual interactions and the Floquet analysis, next-nearest-neighbor hopping terms are introduced into the Floquet Hamiltonian. 
This mechanism, the emergence of next-nearest-neighbor hopping arising from a ``two-step-hopping" process involving nearest-neighbor hopping, is depicted in Fig. \ref{lattice}(d). This mechanism stems from the interactions between higher-order harmonic components and the static component of the Fourier series in the wave function.

As shown in Figs. \ref{lattice}(d), the temporal-periodic hopping coefficient $\lambda_i(t)$ has multiple Fourier components. This time-dependent nature necessitates the eigenmodes to be temporal-periodic, resulting in higher-order harmonics in the wave functions. At the $B$-site, higher-order harmonics in the wave function, such as the first-order harmonic component $\varphi_B^{(1)}$, can interact with the fundamental harmonics $\varphi_A^{(0)}$ and $\varphi_C^{(0)}$ at sites $A$ and $C$, respectively, via the time-periodic interactions $\lambda_i(t)$. Consequently, while lattice sites $A$ and $C$ are not directly connected, the mutual interaction between the fundamental and higher-order harmonics leads to an effective next-nearest-neighbor interaction between $A$ and $C$ from a perturbative perspective.

Temporal-periodic bonds give rise to higher-order harmonic interactions, resulting in the emergence of higher-order harmonic wave functions. As illustrated in Fig. \ref{lattice}(d), these high-order harmonic waves at site $B$ can interact with the zeroth-order wave functions at sites $A$ and $C$. Integrating over these higher-order waves at site $B$ leads to a long-range (next-nearest-neighbor) interaction between sites $A$ and $C$, which were not directly connected in the initial model. This integration process, equivalent to the Floquet-Magnus approximation adopted in Eq. (\ref{floquet_H0_2}), is analogous to the virtual phonon exchange process~\cite{vural2011universal}, in which long-range many-body interactions are established by integrating over the degrees of freedom in virtual phonons.

According to Eqs. (\ref{floquet_H0_3}), it is evident that when the phase delay term $\Delta\phi = 0$ or $\pi$, the time-reversal-symmetry-breaking perturbative Hamiltonian vanishes, and the tight-binding model retains time-reversal symmetry. Furthermore, this perturbative Hamiltonian induces a complex-valued effective hopping interaction with a phase of $\pi/2 + \pi \, {\rm sgn}(\sin \Delta\phi)$. In our model, where $\Delta\phi = \pi/6$, the phase is $\pi/2$. This complex-valued hopping induces a net flux of a time-reversal-breaking gauge field in the area enclosed by the time-modulated interactions, as shown in Fig. \ref{lattice}(e).

Based on the Floquet Hamiltonian shown in Eqs. (\ref{floquet_H0_3}), $H_{\bm{k}}^{\rm NN}$ indicates that the effective hopping between nearest-neighbor sites acquires a complex phase of $\pi/2$, such as the hopping from site $C(\bm{n})$ to site $B(\bm{n})$, as marked by the arrow on the orange solid line in Fig. \ref{lattice}(d). Additionally, $H_{\bm{k}}^{\rm NNN}$ represents the effective next-nearest-neighbor hopping that arises from the time-modulated interactions in the original tight-binding model. It indicates that the next-nearest-neighbor hopping, such as the hopping from site $C(\bm{n})$ to site $A(\bm{n}+(1,0,1))$, as indicated by the arrow on the dashed curved line, has a complex-valued hopping phase of $\pi/2$.

As a result, the phase acquired during the closed-loop hopping $C(\bm{n})\to B(\bm{n})\to A(\bm{n}+(1,0,1))\to C(\bm{n})$ is $\pi/2$. The phase acquired by the closed loop $C(\bm{n})\to A(\bm{n}+(1,0,1))\to B(\bm{n}+(0,1,1))\to C(\bm{n})$ is $3\pi/2$. The phase acquired by the loop $C(\bm{n})\to B(\bm{n})\to C(\bm{n}+(1,0,1))\to B(\bm{n}+(0,1,1))\to C(\bm{n})$ is $2\pi$. The phase acquired by the loop of the hexagon is $\frac{1}{2}\pi\times 6 = 3\pi$. Summarizing these results, we find that each gray-shaded area in Fig. \ref{lattice}(d) has a gauge field flux of $\pi/2$, whereas the blue-shaded area has a flux of zero. The hexagon shown in Fig. \ref{lattice}(e) has non-zero projections in the $x$, $y$, and $z$ directions, leading to an enclosed gauge flux with non-zero projections in all three spatial dimensions. This gauge flux breaks time-reversal symmetry and induces non-zero Chern numbers in all three spatial dimensions, as discussed in Section \uppercase\expandafter{\romannumeral5}.

\begin{figure}
\centering
\includegraphics[width=0.48\textwidth]{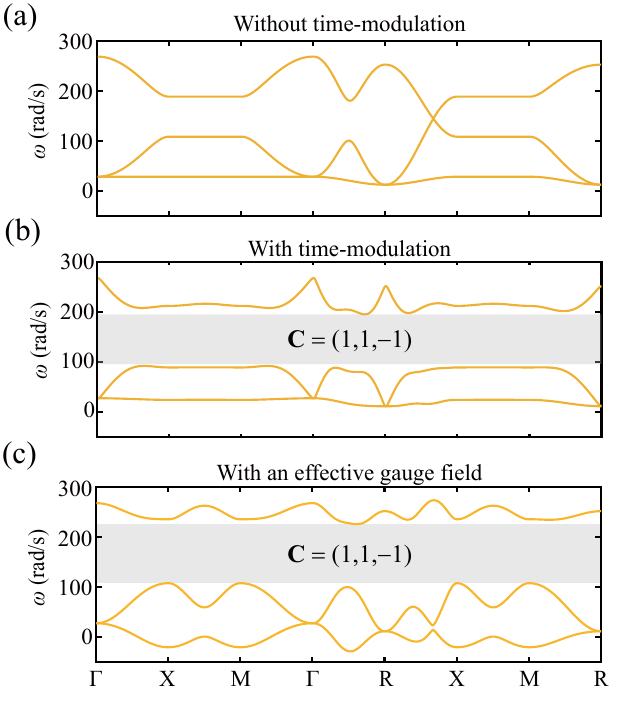}
\caption{The band structure of the tight binding model is considered within the frequency range of $0$ to $300$ rad/s for simplicity. The topological Chern vectors of the bandgap are indicated here. The parameters are given by $m=100$, $\lambda_z=4$, $\omega_F=250$, $\lambda^{(0)}=40$ in (e) and $\lambda^{(1)}=3\lambda^{(0)}=120$ in (f) and (g). (a) The band structure of the tight binding model with all mutual interactions being static in time, meaning the time-modulated hopping coefficient $\lambda^{(1)}$ in Eq. (\ref{Floquet_coup}) is set to $\lambda^{(1)}=0$. (b) The band structure for the truncated Floquet Hamiltonian within the first-order harmonics. This first-order truncated Floquet Hamiltonian is a $9 \times 9$ matrix, which should manifest a total of nine dispersion curves. However, only three out of nine curves fall within the chosen frequency range, and these are presented here. (c) The band structure for the truncated Floquet Hamiltonian within the zeroth-order harmonics, derived using the Floquet-Magnus approximation. The temporal-periodic coefficient is $\lambda^{(1)}=3\lambda^{(0)}=120$ for (b) and (c). The Chern vector for both the first-order and zeroth-order harmonic truncated Floquet Hamiltonian is ${\boldsymbol C} = (1,1,-1)$. 
}\label{band_structure}
\end{figure}

\section{Topological band structure and surface states}

In this section, we study the truncated Floquet Hamiltonian, specifically $\mathcal{H}_{\bm{k}}^{(0)}$ in Eqs. (\ref{floquet_H0}) and (\ref{floquet_H0_2}), to characterize the chiral propagation of topological surface modes. Numerically, we find that achieving a clear topological edge state requires a large band gap, as shown in Fig. \ref{band_structure}. To this end, we set the static part of the mutual interaction to $\lambda^{(0)} = 40$, and the amplitude of the time-modulated part to $\lambda^{(1)} = 3\lambda^{(0)} = 120$. The on-site energy is $m=100$, and the parameter $\lambda_z=4$.

	\begin{figure}[htbp]
		\centering
		\includegraphics[width=0.48\textwidth]{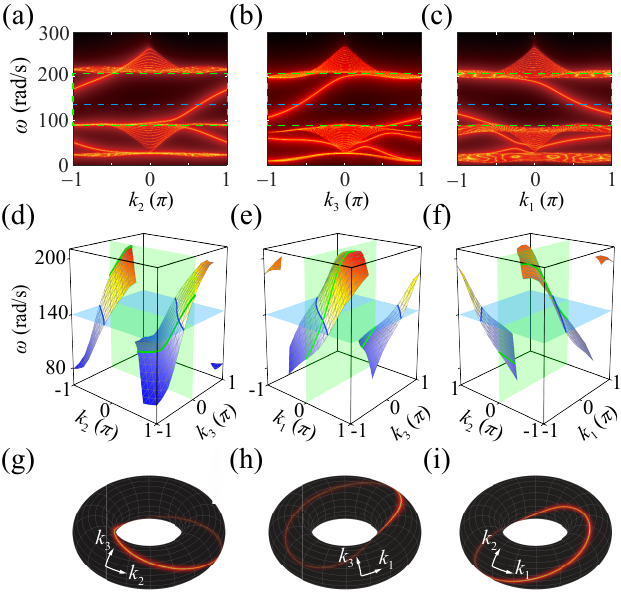}
		\caption{Surface states dispersions of Floquet insulator with a Chern vector $ {\boldsymbol C} = \left( {1,1, - 1} \right)$. (a-c) LDOS of the right(a), front(b) and top(c) surfaces with ${k_3} = 0$ (a-b) and ${k_2} = 0$ (c). (d-f) Dispersion surfaces of surface states. The green intersecting lines between dispersion surfaces and green planes ($k_3 = 0$ in (d-e) and $k_2 = 0$ in (f)) are the dispersions of surface states marked the green dashed boxes in (a-c) respectively. (g-i) LDOS in $\omega  = 140$ plane of different surfaces, showing as the torus knots.\label{supercell}}
	\end{figure}

First, we highlight the importance of breaking time-reversal symmetry by considering the tight binding model in Eq. (\ref{EOM}), with time-modulated hopping interactions set to static. In other words, the time-modulated coefficient in the hopping terms is set to zero, i.e., $\lambda^{(1)}=0$. As a result, the system Hamiltonian $H_{\bm{k}}^{(0)}$ respects time-reversal symmetry. We compute the three dispersion curves of this $3 \times 3$ static Hamiltonian, shown in Fig. \ref{band_structure}(a). The dispersion curves exhibit a gapless band structure due to the presence of time-reversal symmetry, which lacks any non-trivial topology. This result demonstrates the necessity of breaking time-reversal symmetry for achieving Chern class topological insulators, as we show below.

Next, we compute the dispersion relations of $\mathcal{H}_{\bm{k}}^{(1)}$, the truncated Floquet Hamiltonian within the first-order harmonics. This $9 \times 9$ matrix exhibits a total of nine dispersion curves. We plot the middle three dispersion relations (i.e., the fourth, fifth, and sixth dispersions) in Fig. \ref{band_structure}(b). Due to the breaking of time-reversal symmetry, there is a bandgap between the fifth and sixth dispersions, indicating the non-trivial topology of this band structure, as marked by the gray area in Fig. \ref{band_structure}(b). To determine the non-trivial topological index of this bandgap, we compute the Berry curvature of the eigenstates whose frequencies are below the gray bandgap. Specifically, within $\mathcal{H}_{\bm{k}}^{(1)}$, we calculate the Berry curvatures of the lowest five eigenmodes below the bandgap and integrate them over the two-dimensional surface Brillouin zone to find the Chern number for the third dimension. As a result, the Chern vector is given by three components, ${\boldsymbol C}=(C_1, C_2, C_3)$. Each component of the Chern vector is computed as follows: 
\begin{eqnarray}\label{Chern}
C_\alpha(k_\alpha) & = & -\sum_{n={\rm occ}}\int_0^{2\pi}\int_0^{2\pi}\nonumber \\
 & {} & \epsilon_{\alpha\beta\gamma} 2\operatorname{Im}\left\langle  {{\partial }_{{{k}_\beta}}}\psi_n  | {{\partial }_{{{k}_\gamma}}}\psi_n  \right\rangle d{{k}_\beta} d{{k}_\gamma}. 
\end{eqnarray}
Here, $\sum_{n={\rm occ}}$ denotes the occupied bands below the bandgap, and $\alpha, \beta, \gamma = 1, 2, 3$ label the three wave numbers $k_1$, $k_2$, and $k_3$. In principle, each component $C_\alpha(k_\alpha)$ of the Chern vector depends on the choice of the wavenumber $k_\alpha$ and therefore can jump from integer to integer. However, in the current model that we study, which has a gapped band structure spanning the entire Brillouin zone, these Chern vector components are well-defined and do not change as $k_\alpha$ varies, allowing us to denote them as $C_\alpha(k_\alpha) = C_\alpha$. Using the parameters shown in this section, we obtain the Chern vector as ${\boldsymbol C}=(1, 1, -1)$, indicating the topologically non-trivial Chern class of the Floquet model within the first-order harmonic truncation due to the broken time-reversal symmetry.

Finally, we approximate the Hamiltonian within the zeroth-order harmonics. The resultant Hamiltonian, $\mathcal{H}_{\bm{k}}^{(0)}$, is represented in Eq. (\ref{floquet_H0_2}). We compute the three dispersion curves for this $3 \times 3$ matrix, as shown in Fig. \ref{band_structure}(c). The first and second dispersion curves exhibit a bandgap that spans the entire Brillouin zone compared to the third dispersion curve. We then compute the Berry curvature and Chern vector for the lowest two dispersion curves, resulting in ${\boldsymbol C} = (1, 1, -1)$. Comparing the results in Figs. \ref{band_structure}(b) and (c), it is evident that although the band structures differ, they possess the same topological numbers. This indicates the robustness of these topological numbers against perturbations in the Hamiltonian.

\begin{figure}[htbp]
		\centering
		\includegraphics[width=0.47\textwidth]{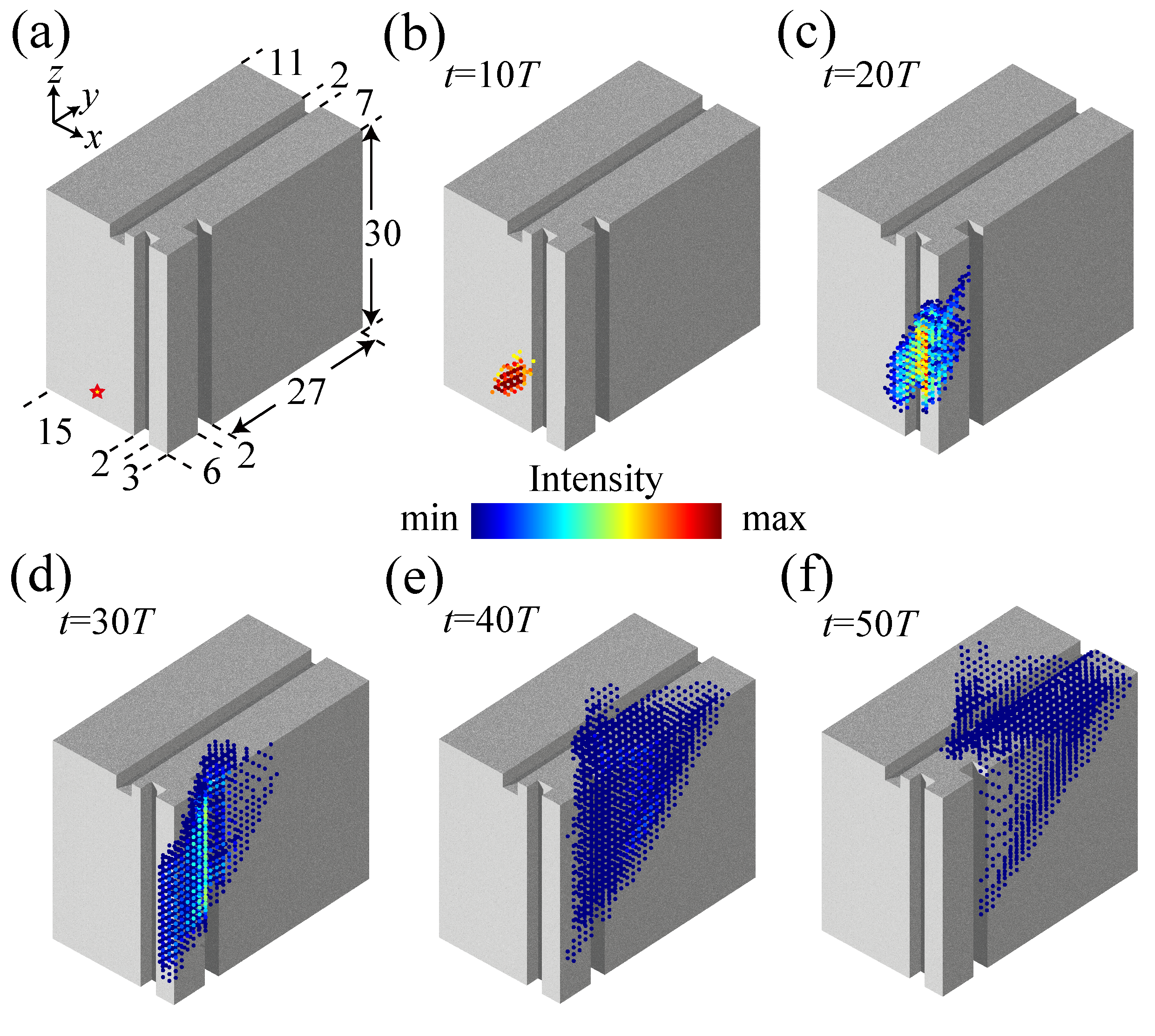}
		\caption{Numerically simulated propagation of a Floquet surface state in a finite lattice with defects. (a) The specific shape of the lattice. Yellow star denotes the point source. (b-f) The intensity distributions of the excited surface states at different times.\label{simulation}}
	\end{figure}

Having established the topological band structure, we now investigate whether topologically protected surface states can arise in this three-dimensional lattice. To this end, we perform a supercell analysis of the time-modulated stacked kagome lattice. We construct the lattice with $N_1\times N_2\times N_3$ ($N_1=20$, $N_2=35$, $N_3=30$) unit cells along the $\bm{a}_1$, $\bm{a}_2$, and $\bm{a}_3$ primitive vector directions. Figure \ref{simulation} presents the numerical observation of these chiral surface states. Specifically, this supercell analysis is based on the established lattice, as shown in Fig. \ref{simulation}, incorporating three surface defects parallel to the $\bm{a}_1$, $\bm{a}_2$, and $\bm{a}_3$ primitive vector directions. In Figs. \ref{simulation}(a), (b), and (c), the structures are subjected to open boundary that is normal to the reciprocal vector $\bm{b}_1$ direction (with Bloch boundary conditions in the $\bm{a}_2$ and $\bm{a}_3$ directions), open boundary conditions in the $\bm{b}_2$ direction (with Bloch boundary conditions in the $\bm{a}_3$ and $\bm{a}_1$ directions), and open boundary conditions in the $\bm{b}_3$ direction (with Bloch boundary conditions in the $\bm{a}_1$ and $\bm{a}_2$ directions), respectively. We then analyze the dispersions of the right, front, and top surfaces, with normal vectors $\bm{b}_1$, $\bm{b}_2$, and $\bm{b}_3$, respectively.

The supercell band structures are shown in Fig. \ref{supercell}(a), (b), and (c). The band structure is weighted by the spatial distribution of the wave function around the lattice boundary, which is included in the numerator of the following definition:
\begin{eqnarray}
g\left( \omega, k_\beta, k_\gamma \right)=\sum\limits_{i}{\frac{\sum\limits_{j\in {\rm surface \,\,normal \,\, to \,\,}\bm{b}_\alpha}{{{\left| \psi _i^{(j)}(k_\beta, k_\gamma)\right|}^{2}}}}{\pi [ {{\left( \omega -{\omega }_i \right)}^{2}}+{\Gamma^2} ]}\Gamma }\nonumber \\ 
\end{eqnarray} 
where $\alpha, \beta, \gamma=1,2,3$ denote the three components of the wavenumbers, $\omega_i$ is $i$th eigenvalue of supercell Hamiltonian, $\sum_i$ sums over all eigenvalues in the supercell band structure, $\omega$ is the frequency in the band structure, and $\Gamma=0.01$ rad/s represents the width of the dispersion curves. $\sum_j$ sums over all sites on the open surface that is normal to the reciprocal vector $\bm{b}_\alpha$, and $\psi_i^{(j)}$ denotes the wave function component on the $j$th site of the $i$th eigenstate, whose wavenumbers are $k_\beta$ and $k_\gamma$ in the two periodic boundaries.

As shown in Figs. \ref{supercell}(a-c), the dispersion curves of the surface states exhibit strong asymmetry due to the breaking of time-reversal symmetry. In Figs. \ref{supercell}(a) and (b), the in-gap surface states exhibit positive group velocities, reflecting the Chern numbers $C_1 = C_2 = 1$. In contrast, the surface state in Fig. \ref{supercell}(c) has a negative group velocity, corresponding to the Chern number $C_3 = -1$ in the topological Chern vector. These chiral surface states can propagate without backscattering along the surfaces, as it is impossible to acquire the opposite group velocity according to the in-gap dispersion relations.  

To better visualize the dispersion relations of the chiral surface states, we plot them in Figs. \ref{supercell}(d-f) with respect to the surface Brillouin zone. Specifically, the dispersion relations for these states are plotted with respect to the surface Brillouin zones of $(k_2, k_3)$, $(k_3, k_1)$, and $(k_1, k_2)$. Due to the topologically protected nature of all three components in the Chern vector, the sign of the group velocity of these chiral surface states is fixed and does not change as the wavenumber in the surface Brillouin zones varies. This is a manifestation of the topologically protected chirality in this Chern vector topological insulator.

Next, we provide a mathematical description of each component of the Chern vector. We study the Fermi energy at the frequency $\omega = 140$ rad/s in Fig. \ref{supercell}(d-f). The intersection between the in-gap dispersion relations and the Fermi energy is known as the surface Fermi loops~\cite{vanderbilt2018berry}. As shown in Fig. \ref{supercell}(g-i), these Fermi loops form knots in the Brillouin zone torus. The number of times a knot winds around the longitude and meridian of the torus reflects the component of the Floquet Chern vector along the corresponding direction. For example, Fig. \ref{supercell}(g) represents a $\left( C_2=1, C_3=-1 \right)$ torus knot, where the winding numbers of the surface Fermi loop indicate the Chern number of $C_2=1$ along the wavenumber $k_2$ and $C_3=-1$ along $k_3$.

In contrast to the trivial stacking of two-dimensional Chern insulators in the third dimension, where chiral surface states can only propagate in the horizontal directions, this three-dimensional Chern vector insulator possesses chiral surface states in all three spatial dimensions. We study the propagation and robustness of surface states on the surfaces of the lattice in the presence of structural defects. To this end, we establish a lattice subjected to open boundary conditions on all surfaces that are normal to the $\bm{b}_1$, $\bm{b}_2$, and $\bm{b}_3$ directions. As mentioned earlier, the most remarkable property of the Chern vector topological insulator is its topologically protected unidirectional surface states. To confirm this, we consider a finite lattice with open boundary conditions in all directions, constituted by the unit cell shown in Fig. \ref{lattice}(b). The lattice further possesses three defects located on the right, front, and top surfaces, respectively, as shown in Fig. \ref{simulation}(a).

We numerically simulate the propagation of a surface state based on the original time-dependent Hamiltonian $H(t)$ shown in Eq. (\ref{real_space_TB}). The lattice is excited using a pulse signal from a point source (marked by a star) with a center frequency of $\omega = 140$ rad/s. In Fig. \ref{simulation}(b), the resulting signal forms a small two-dimensional wave packet on the front surface at $t = 10T$ ($T=2\pi/\omega$) without leaking into the interior of the lattice. As $t$ increases from $10T$ to $50T$, the wave packet moves directionally along the surfaces as a whole, gradually spreading due to the distribution of group velocity. This directional movement exhibits strong non-reciprocity of signal propagation, indicating that the excited surface states are chiral on every surface of this lattice. Additionally, the wave packet bypasses all three defects without backscattering, demonstrating strong topological protection against structural defects.

\section{Conclusion}
In this work, we have proposed a three-dimensional temporal-periodic tight binding model that effectively breaks time-reversal symmetry. The truncated Floquet Hamiltonian manifests a time-reversal-breaking gauge field, inducing a Chern vector with three components that are topologically protected Chern numbers. Unlike the trivial stacking of two-dimensional Chern insulators in the third dimension, where chiral edge states only propagate horizontally, this three-dimensional Chern vector can host edge states that move on all surfaces of the structure. These topological surface states have been numerically analyzed, exhibiting chirality and strong topological protection against structural defects on all surfaces of the 3D insulator.

While breaking time-reversal symmetry is challenging for classical systems such as mechanical~\cite{lei2022duality, Ma2023PRL}, elastic~\cite{chen2021research, yang2023non, martello2023coexistence}, acoustic~\cite{chen2023various, rosa2023material}, and electric circuit~\cite{hu2024observation2} systems, our work proposes an alternative route using time-modulated elements to effectively induce a time-reversal-breaking gauge field in the model. Future research can focus on experimentally realizing three-dimensional Chern vector topological insulators in classical metamaterials~\cite{PhysRevLett.130.017201, wen2019acoustic, long2020realization, Zhou2018PRL}, extending the Floquet technique to systems with higher Chern numbers~\cite{bosnar2023high}, and exploring hyperbolic Chern insulators~\cite{PhysRevLett.129.088002, PhysRevB.105.245301, PhysRevB.110.195113}.


 \section{Acknowledgment}
Di Zhou wishes to acknowledge the insightful discussions with Dr. Xueda Wen. The authors wish to acknowledge the support from the National Nature Science Foundation of China (Grant Nos. 12374157, 12074446, 12102039, and 12302112), and the Fundamental Research Funds for the Central Universities (No. 30923010207).


%

\end{document}